\documentclass[useAMS,usenatbib]{mn2e}
\usepackage{graphicx}

  \newcommand{\ParDiff}[2]{\frac{\partial{#1}}{\partial{#2}}}
  \newcommand{\Bracket}[1]{{\left({#1}\right)}}
  
  \newcommand{\MathVector}{\mathbf}

  \newcommand{\BH}{black hole}

\newcommand{\Degree}{{{}^\circ}}

\newcommand{\AM}{angular momentum}
\newcommand{\VC}{viscosity coefficient}

\newcommand{\xUnitVec}{{\MathVector{e}_x}}
\newcommand{\yUnitVec}{{\MathVector{e}_y}}
\newcommand{\zUnitVec}{{\MathVector{e}_z}}
\newcommand{\lxy}{{\lVec_{xy}}}

\newcommand{\La}{{\LaVec}}
\newcommand{\Ladot}{{\LadotVec}}
\newcommand{\Lsurf}{{\LsurfVec}}
\newcommand{\Lsurfdot}{{\LsurfdotVec}}
\newcommand{\Ls}{\LsVec}

\newcommand{\OmegaPre} {\OmegaPreVec}
\newcommand{\omegaPre} {\omegaPreVec}
\newcommand{\Tvis} {\TvisVec}

\newcommand{\SigmaA}{{\Sigma_a}}
\newcommand{\OmegaK}{{\Omega_K}}

\newcommand{\lVec} {{\MathVector{l}}}
\newcommand{\LVec} {{\MathVector{L}}}
\newcommand{\LaVec} {{\MathVector{L}_a}}
\newcommand{\LadotVec} {{\MathVector{\dot{L}}_a}{}}
\newcommand{\LadotSca} {{{\dot{L}}_a}{}}
\newcommand{\LsurfVec} {{\MathVector{L}_{surf}}}
\newcommand{\LsurfdotVec} {{\MathVector{\dot{L}}_{surf}}{}}
\newcommand{\LsVec} {{\MathVector{L}_s}}

\newcommand{\OmegaVec} {{\MathVector{\Omega}}}
\newcommand{\OmegaPreVec} {{\MathVector{\Omega}_{pre}}}
\newcommand{\omegaPreVec} {{\MathVector{\omega}_{p}}}
\newcommand{\TvisVec} {{\MathVector{T}_{vis}}}

\newcommand{\LaSca}{{L_a}}

\newcommand{\OmegaPreSca} {{\Omega_{pre}}}
\newcommand{\omegaPreSca} {{\omega_{p}}}

\newcommand{\Vr}{{V_r}}
\newcommand{\Vx}{{V_x}}

\newcommand{\AMBH} {\MathVector{J}_H}

\newcommand{\Rw}{R_{\rm w}}

\title{Steady state solution of warped accretion discs}
\author
[Lei Chen et al.]{Lei Chen$^1$%
\thanks{Email: lchen@shao.ac.cn},
Shengmiao Wu$^1$%
\thanks{Email: smwu@shao.ac.cn}
and Feng Yuan$^1$%
\thanks{Email: fyuan@shao.ac.cn}
\\$^1$
Key Laboratory for Research in Galaxies and Cosmology,
Shanghai Astronomical Observatory, Chinese Academy of Sciences,
\\80 Nandan Road, Shanghai 200030, China
}

\begin{document}

\maketitle

\defcitealias{Martin2007}{MPT07}
\defcitealias{Martin2008}{M08}
\defcitealias{Scheuer1996}{SF96}

\begin{abstract}
{ We consider a thin accretion disc warped due to the
Bardeen-Petterson effect, presenting both analytical and numerical
solutions for the situation that the two \VC{s} vary with radius as
power law, with the two power law indices not necessarily equal.
The analytical solutions are compared with numerical ones, showing that
our new analytical solution is more accurate than previous one,
which overestimates the inclination changing in the outer disc.
Our new analytical solution is appropriate for moderately warped
discs, while for extremely misaligned disc, only numerical solution
is appropriate. }
\end{abstract}

\begin{keywords}
accretion discs -- black hole physics -- galaxies: nuclei
\end{keywords}

\section{Introduction}
Observational evidences are accumulating that accretion discs
around \BH{s} can be warped.
Warped accretion discs have been directly observed
by water maser observations in NGC4258 
\citep{Miyoshi1995, Neufeld1995, Herrnstein1996}
and Circinius galaxy
\citep{Greenhill2003}.
The lack of correlation of radio jets in AGNs
and the disc plane of host galaxy \citep{Kinney2000, Schmitt2002}
can also be explained by disc warping.
\citet{Wu2008}
discussed the possibility that double-peaked Balmer lines
in AGNs be emitted by warped disc.
Possible evidence for disc warping is also found in X-ray binaries,
including the misalignment between jets and orbital planes in GRO J 1655-40
\citep{Greene2001, Hjellming1995},
and the precessing of jets in SS433 \citep{Blundell2004}.

Theoretically, warpping can be caused by various mechanisms, including
tidally induced warping by a companion in a binary system
\citep{Terquem1993, Larwood1996, Terquem1996},
radiation driven or self-inducing warping,
\citep{Maloney1996,Maloney1997,Maloney1998, Pringle1996,Pringle1997},
magnetically driven disc warping,
\citep{Lai1999, Lai2003, Pfeiffer2004},
and frame dragging driven warping \citep{Bardeen1975}.
Herein we consider the shape of a disc warped by the last mechanism.

\citet{Bardeen1975} pointed out that,
the combining effect of Lense-Thirring effect
and the viscosity within the disc cause the inner part of the disc
to be aligned with the central \BH,
while the outer part of disc remains tilted,
thus resulting in a warped disc.
\citet{Pringle1992} derived the dynamical equations of such a warped disc.
\citet[][hereafter \citetalias{Scheuer1996}]{Scheuer1996} analytically solved the equation with a first order approximation,
assuming constant \VC{s}.
\citet{Lodato2006} numerically solved the equations, also assuming constant \VC{s}.
\citet[][hereafter \citetalias{Martin2007}]{Martin2007}
generalized \citetalias{Scheuer1996}'s analytical solution to the situation
that the \VC{s} varies as power law,
and then, \citet[][hereafter \citetalias{Martin2008}]{Martin2008}
used this solution to fit the maser observation
of NGC4258's disc.

We carried on a numerical calculation for a warped disc
with power-law varying $\nu$,
and compared the results with \citetalias{Martin2007}'s analytical solution.
The importance of this work lies in such a fact:
\citetalias{Martin2007}'s analytical solution
(and \citetalias{Scheuer1996}'s, as well)
are based on first order approximation,
under the assumption of a small inclination angle $\theta_{out}\ll1$,
while the real accretion discs can be strongly misaligned $\theta_{out}\sim1$,
e.g., the fitting of NGC4258 shows a strong misaligning.
A numerical calculation is needed to tell exactly how the error grows.
Our calculation shows a prominent deviation between
analytical solution and exact solution when the inclination angle are large,
suggesting that the analytical solutions not appropriate for
study of NGC4258 or other strongly misaligned discs.

We then proposed another way to extrapolate the small $\theta_{out}$ solution
to large $\theta_{out}$ situation,
and thus find a new analytical solution.
The new solution is also compared with numerical calculation
and proves to be more accurate for large $\theta_{out}$ situation.
We also generalized the analytical solutions
to the situation
that $\nu_1$ and $\nu_2$ have different power index.

\section{The basic scenario and equations}
We use the assumptions the same as adopted by \citet{Pringle1992}.
The disc is assumed to be a thin one,
consisting of concentric (but misaligned) circular gas rings.
Each ring can be totally described with its surface density $\Sigma$,
its angular velocity $\OmegaVec$,
and its radial velocity $\Vr$.
Note that $\OmegaVec$ is a vector,
so that it describes both the speed of the rotation $\Omega=|\OmegaVec|$
and the orientation of the ring $\lVec=\OmegaVec/\Omega$.
So, the state of the disc can be totally described by
the distribution of the three quantities with radius $R$,
$\Sigma=\Sigma(R)$, $\OmegaVec=\OmegaVec(R)$,
and $\Vr=\Vr(R)$.
Each ring will receive viscous torque from neighbouring rings
whenever the angular velocity $\OmegaVec$ changes with radius,
$\ParDiff{\OmegaVec}{R}\neq0$.
Each ring also receive a Lense-Thirring torque from the central \BH{}
whenever it is misaligned with the \BH.
The dynamical equations under such assumption are
\begin{equation}
\begin{array}{lcl}
  \dot\Sigma & = &
	-\frac{1}{R} \Bracket{R \Sigma \Vr}'
  \\
  {\Lsurfdot} & = &
	-\frac{1}{R}\Bracket{R \Vr \Lsurf}' + \frac{1}{R}\Tvis' + \OmegaPre\times\Lsurf
  \\
  \Tvis &=&
	R^3 \Sigma \Bracket
	{ \nu_1\Omega'\lVec + \frac{\nu_2}{2}\Omega\lVec' }
  \\
  \Omega &=&
	\OmegaK
\end{array}
\end{equation}
Where $\OmegaPre$ is the Lense-Thirring precession frequency
\begin{equation}
  \OmegaPre=\omegaPre/R^3=\frac{2G\AMBH}{c^2R^3}
\end{equation}
$\Lsurf=\Sigma\Ls$ is the surface density of \AM.
$\Ls=R^2\OmegaVec$ is the specific \AM,
i.e., the \AM{} carried by unit mass.
Here we use a dot on the head to stand for $\ParDiff{}{t}$,
and the prime symbol ``$'$'' to stand for $\ParDiff{}{R}$.

In this work, we use logarithemic coordinate $x=\ln(R/R_0)$,
where $R_0$ is an arbitrarily defined length scale,
so that all the physical quantities shall be written as functions of $x$.
The mass of a ring $x\sim x+dx$ is $dm=\Sigma\cdot2\pi RdR=2\pi\SigmaA dx$,
where annulus density $\SigmaA=R^2\Sigma$ is the mass on unit $x$ interval and unit arc angle.
The angular momentum of the ring is
$d\LVec=2\pi\La dx=2\pi\SigmaA\Ls dx$,
where annulus \AM{} density $\La=\SigmaA\Ls$
is the \AM{} on unit $x$ interval and unit arc angle.
And we describe the radial motion of rings with $\Vx=\frac1R \Vr$,
which is the $x$ interval the ring moves in unit time.
So the disc can be describe with ($\SigmaA$, $\La$, $\Vx$),
as functions of $x$,
and the evolution of the disc is described
with the evolution of the functions with time $t$.
In the following we use a dot on the head to stand for $\ParDiff{}{t}$,
and the prime symbol ``$'$'' to stand for $\ParDiff{}{x}$.

With the denotation defined above,
the equations can be written in a simpler form
(nevertheless equivalent to the previous form).
\begin{equation}\label{Equation:Basics}
\begin{array}{lcl}
  \dot\SigmaA & = &
	-\Bracket{\SigmaA \Vx}'
  \\
  {\Ladot} & = &
	\Bracket{{\Ladot}}_{adv} + \Bracket{{\Ladot}}_{vis}
	+ \Bracket{{\Ladot}}_{pre}
  \\
  & = &
	-\Bracket{\Vx \La}' + \Tvis' + \OmegaPre\times\La
  \\
  \Tvis &=&
	\SigmaA \Bracket
	{ \nu_1\Omega'\lVec + \frac{\nu_2}{2}\Omega\lVec' }
  \\
  \Omega &=&
	\OmegaK
\end{array}
\end{equation}
Note that the ``$'$'' here means $\ParDiff{}{x}$ instead of $\ParDiff{}{R}$,
and $\ParDiff{}{x}=R\ParDiff{}{R}$.

\section{Steady state solution for slightly misaligned disc}
Under Keplerian assumption, the disc can be entirely depicted
by a distribution of $\La$.
Eliminating redundant variables, 
eqs.(\ref{Equation:Basics}) can be rewritten as
\begin{equation}\label{Equation:AM}
\begin{array}{lcl}
  {\Ladot} & = & \displaystyle
	- \Bracket{\frac32\nu_1\frac{1}{R^2}\La}'
	+ \Bracket{\frac{\nu_2}{2}\frac{1}{R^2}\LaSca\lVec'}'
  \\ & & \displaystyle
	+ \Bracket{3\Bracket{\nu_1\frac{1}{R^2}\LaSca}'\lVec}'
	+ \Bracket{\nu_2\frac{1}{R^2}\Bracket{\lVec'}^2\La}'
  \\ & & \displaystyle
	+ \OmegaPre\times\La
\end{array}
\end{equation}
By $\lVec\cdot$eq.(\ref{Equation:AM}), we get the parallel part of the equation.
\begin{equation}\label{Equation:AMparaPart}
\begin{array}{lcl}
  {\LadotSca} & = & \displaystyle
	- \Bracket{\frac32\nu_1\frac{1}{R^2}\LaSca}'
	- \Bracket{\frac{\nu_2}{2}\frac{1}{R^2}\LaSca}\Bracket{\lVec'}^2
  \\ & & \displaystyle
	+ 3\Bracket{\nu_1\frac{1}{R^2}\LaSca}''
	+ \Bracket{\nu_2\frac{1}{R^2}\Bracket{\lVec'}^2\LaSca}'
\end{array}
\end{equation}
By eq.(\ref{Equation:AM})$-\lVec\times$eq.(\ref{Equation:AMparaPart}),
we get the perpendicular part of the equation.
\begin{equation}\label{Equation:AMperpPart}
\begin{array}{lcl}
  {\LaSca}\dot\lVec & = & \displaystyle
	- \Bracket{\frac32\nu_1\frac{1}{R^2}\LaSca}\lVec'
	+ \Bracket{\frac{\nu_2}{2}\frac{1}{R^2}\LaSca}'\lVec'
  \\ & & \displaystyle
	+ \Bracket{\frac{\nu_2}{2}\frac{1}{R^2}\LaSca}\Bracket{\lVec''}_\perp
	+ 3\Bracket{\nu_1\frac{1}{R^2}\LaSca}'\lVec'
  \\ & & \displaystyle
	+ \Bracket{\nu_2\frac{1}{R^2}\Bracket{\lVec'}^2\LaSca}\lVec'
	+ \OmegaPre\times\La
\end{array}
\end{equation}

When the disc is only slightly misaligned,
eqs(\ref{Equation:AM}) can be linearized.
Taking the z-axis along the direction of $\OmegaPre$,
we have
$\lVec=l_x\xUnitVec+l_y\yUnitVec+l_z\zUnitVec
\approx \zUnitVec + \lxy$,
where $\lxy=l_x\xUnitVec+l_y\yUnitVec$,
when $l_x$ and $l_y$ are small enough
for their second-order term to be neglected.
Then $\lVec'=\lxy'=l_x'\xUnitVec+l_y'\yUnitVec$,
$\lVec''=\lxy''=l_x''\xUnitVec+l_y''\yUnitVec$,
and $\lVec''\cdot\lVec=(\lVec')^2=0$.
Using these approximations, 
the two parts of the \AM{} equations becomes
\begin{equation}\label{Equation:Linearized0}
\begin{array}{lcl}
\LadotSca &=& \displaystyle
	-\Bracket{\frac32\nu_1\frac{1}{R^2}\LaSca}'
	+\Bracket{3\nu_1\frac{1}{R^2}\LaSca}''
\\
L_a \dot{\lxy} &=& \displaystyle
	-\frac32\nu_1\frac{1}{R^2}\LaSca \lxy'
	+ \Bracket{\frac{\nu_2}{2}\frac{1}{R^2}\LaSca \lxy'}'
\\ & &
	+ 3\Bracket{\nu_1\frac{1}{R^2}\LaSca}'\lxy'
	+ \OmegaPre \times \Bracket{\LaSca \lxy}
\end{array}
\end{equation}
Further using SF96 and MPT07's symbol $W=l_x+il_y$, where $i=\sqrt{-1}$,
the equation become
\begin{equation}\label{Equation:Linearized}
\begin{array}{lcl}
\LadotSca &=& \displaystyle
	-\Bracket{\frac32\nu_1\frac{1}{R^2}\LaSca}'
	+\Bracket{3\nu_1\frac{1}{R^2}\LaSca}''
\\
L_a \dot{W} &=& \displaystyle
	-\frac32\nu_1\frac{1}{R^2}\LaSca W'
	+ \Bracket{\frac{\nu_2}{2}\frac{1}{R^2}\LaSca W'}'
\\ & & \displaystyle
	+ 3\Bracket{\nu_1\frac{1}{R^2}\LaSca}'W'
	+ i\OmegaPreSca \LaSca W
\end{array}
\end{equation}
It is not surprising that the first part is all the same with
that for a planary disc.
This means for slightly warped disc we can find the solution in two steps.
In first step the evolution and distribution of $\LaSca$ are solved,
with the misaligning omitted and the disc looked upon as planary.
In second step the inclination at each radius are found,
with $\LaSca$ already known.
This two-step method is much easier than finding the exact solution.

To find a steady state solution,
we set the left side of eqs.(\ref{Equation:Linearized}) to zero
\begin{equation}\label{Equation:LinearizedSteadyState}
\begin{array}{lcl}
0 &=& \displaystyle
	-\Bracket{\frac32\nu_1\frac{1}{R^2}\LaSca}'
	+\Bracket{3\nu_1\frac{1}{R^2}\LaSca}''
\\
0 &=& \displaystyle
	-\frac32\nu_1\frac{1}{R^2}\LaSca W'
	+ \Bracket{\frac{\nu_2}{2}\frac{1}{R^2}\LaSca W'}'
\\ & & \displaystyle
	+ 3\Bracket{\nu_1\frac{1}{R^2}\LaSca}'W'
	+ i\OmegaPreSca \LaSca W
\end{array}
\end{equation}
The solution of $\LaSca$ are simple
\begin{equation}\label{Equation::LaScaSolution}
\nu_1\LaSca= C_0R^{5/2}+C_1R^2
\end{equation}
where $C_0$ and $C_1$ are constants.
$C_1$ is connected with the condition at inner boundary,
and always become unimportant when the concerned region
are much larger than inner radius.
So we discard $C_1$ and get
\begin{equation}
\nu_1\LaSca= C_0R^{5/2}
\end{equation}
Substituting the $\LaSca$ value back, we get
\begin{equation}\label{Equation:W1}
0 =
	\Bracket{\frac{\nu_2}{2\nu_1}R^{1/2} W'}'
	+ \frac{i\omegaPreSca}{\nu_1} R^{-1/2} W
\end{equation}
If $\nu_1$ and $\nu_2$ vary with radius as power law
$\nu_1=\nu_{10} (R/R_0)^{\beta_1}= \nu_{10} \exp(\beta_1 x)$,
$\nu_2= \nu_{20} \exp(\beta_2 x)$, 
the equation of $W$ becomes
\begin{equation}\label{Equation:W2}
\begin{array}{l}
  \displaystyle
	\Bracket{\frac{\nu_{20}R_0}{2}\exp[(1/2+\beta_2-\beta_1)x] W'}'
  \\ \displaystyle
  + i\omegaPreSca \exp[(-1/2-\beta_1)x] W
	= 0
\end{array}
\end{equation}
Physically we have the boundary conditions
\begin{equation}
\begin{array}{lll}
W\rightarrow0		&	R\rightarrow0
\\
W\rightarrow W_\infty	&	R\rightarrow\infty
\end{array}
\end{equation}
Solving eq.(\ref{Equation:W2}) under such boundary conditions,
we get
\begin{equation}\label{Equation:AnalyticalSolutionA}
W= f W_\infty 
\end{equation}
where
$$
f= \frac{2^{1-n}}{\Gamma(n)} s^n K_n(s)
$$
and
$$n=\frac{1/2+\beta_2-\beta_1}{1+\beta_2}$$
and
$$
s = \frac{2}{1+\beta_2}(1-i)\sqrt{\frac{\omegaPreSca}{\nu_{20}R_0}}
		\exp\Bracket{-\frac{1+\beta_2}{2}x}
$$
and $K_n$ is the $n$th order modified Bessel function of the second kind.
The solution reduces to the MPT07 one (see eq.(24) therein)
when the two \VC{s} vary with same index
$\beta_1=\beta_2=\beta$,
and further to SF96 solution when $\beta_1=\beta_2=0$.

By defining the warping radius as
\begin{equation}
\Rw=\Bracket{ \frac{\omegaPreSca}{\nu_{20}}R_0^{\beta_2} } ^{1/(1+\beta_2)}
\end{equation}
the parameter $s$ can be written as
\begin{equation}
s = \frac{2}{1+\beta_2}(1-i)\Bracket{\frac{R}{\Rw}}
		^{-\frac{1+\beta_2}{2}}
\end{equation}
Hereafter we always set $R_0=\Rw$, i.e.,
use the warping radius as length unit,
thus making the problem scale-free,
and turning the equation into
\begin{equation}
s = \frac{2}{1+\beta_2}(1-i)
		\exp\Bracket{-\frac{1+\beta_2}{2}x}
\end{equation}
It is easy to see that the warping radius thus defined
is where the Lense-Thirring precessing timescale
and viscosity timescale equals
\begin{equation}
\frac{1}{\OmegaPreSca(\Rw)}\equiv\frac{\Rw^3}{\omegaPreSca}
=\frac{\Rw^2}{\nu_2(\Rw)}
\end{equation}

We present here the analytical solutions
for several sets of $\beta_1$ and $\beta_2$.
The $\beta_1$ values are 0, 1, 2, respectively,
and for each $\beta_1$ we calculated for $\beta_2=\beta_1$,
$\beta_2=\beta_1+0.1$, $\beta_2=\beta_1-0.1$.
The plane of $z$ axis and $\lVec$ at infinite radius, $\lVec_{out}$,
is set to be the $xz$ plane,
so that $\lVec_{out}=(\sin\theta_{out},0,\cos\theta_{out})$ 
and $W_\infty=\sin\theta_{out}$.
For each solution we plot in Fig.1 the $f$ value in the complex plane,
which is equivalent to an $l_y/W_\infty$ against $l_x/W_\infty$ plot.
In Fig.2 we plot the absolute value and angle of $f$ (divided by $2\pi$) against radius $x$.
The angle of $f$ equals the azimuthal angle $\varphi$ of $\lVec$.
The absolute value of $f$ is 
$|f|=\frac{|W|}{W_\infty}=\frac{\sin\theta}{\sin\theta_{out}}$,
and equals $\frac{\theta}{\theta_{out}}$ for small $\theta_{out}$.
So Fig.2 is also $\frac{\theta}{\theta_{out}}$ against $x$
and $\frac{\varphi}{2\pi}$ against $x$ plots.
We divide $\varphi$ by $2\pi$ so that
the value is now the turns $\lVec$ has precessed around $z$ axis.
The very fast growth of $\varphi$ in the innermost part of disc
is not important, because $\theta$ is already very small there,
meaning the disk is almost aligned will \BH{} spin.

In the following we call
eq.(\ref{Equation:AnalyticalSolutionA}) ``solution A".

\begin{figure}
\includegraphics[angle=-90,scale=0.3]{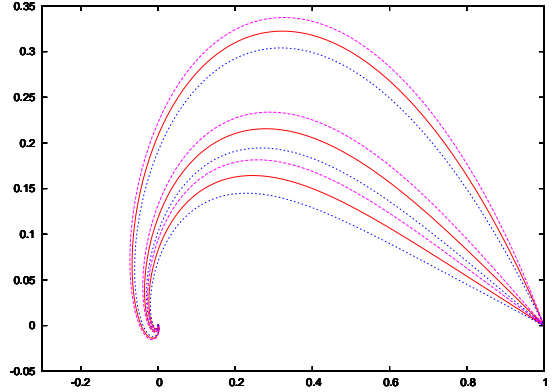}
\caption{Analytical solutions, $l_y/W_\infty$ against $l_x/W_\infty$.
	The solid lines: $\beta_2=\beta_1$;
	the long dash lines: $\beta_2=\beta_1+0.1$;
	the short dash lines: $\beta_2=\beta_1-0.1$.
	For each line style the three lines are for
	$\beta_1=$0, 1, 2, respectively, from upside to downside.
	}
\end{figure}

\begin{figure}
\center
\includegraphics[angle=-90,scale=0.3]{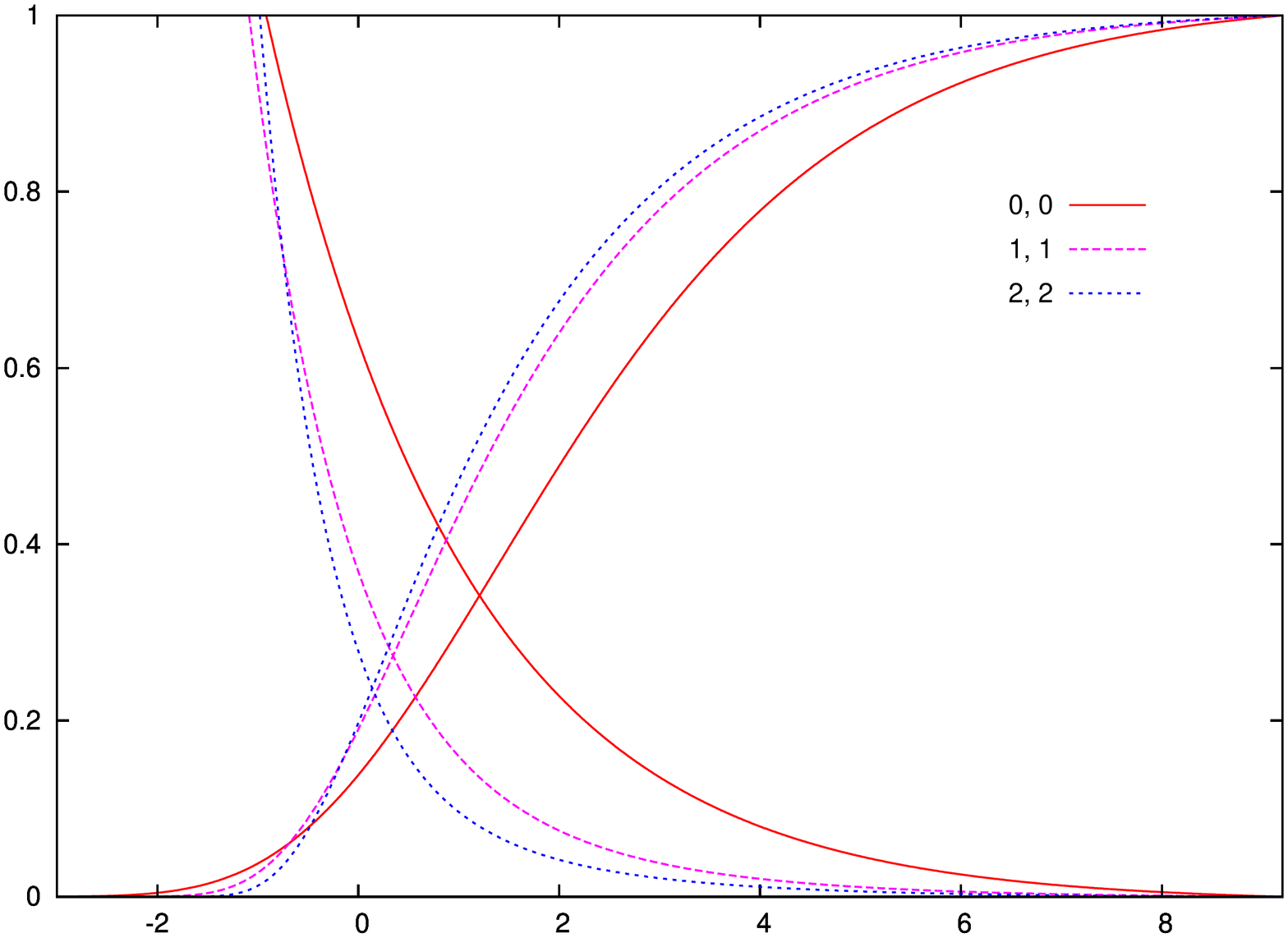}
\includegraphics[angle=-90,scale=0.3]{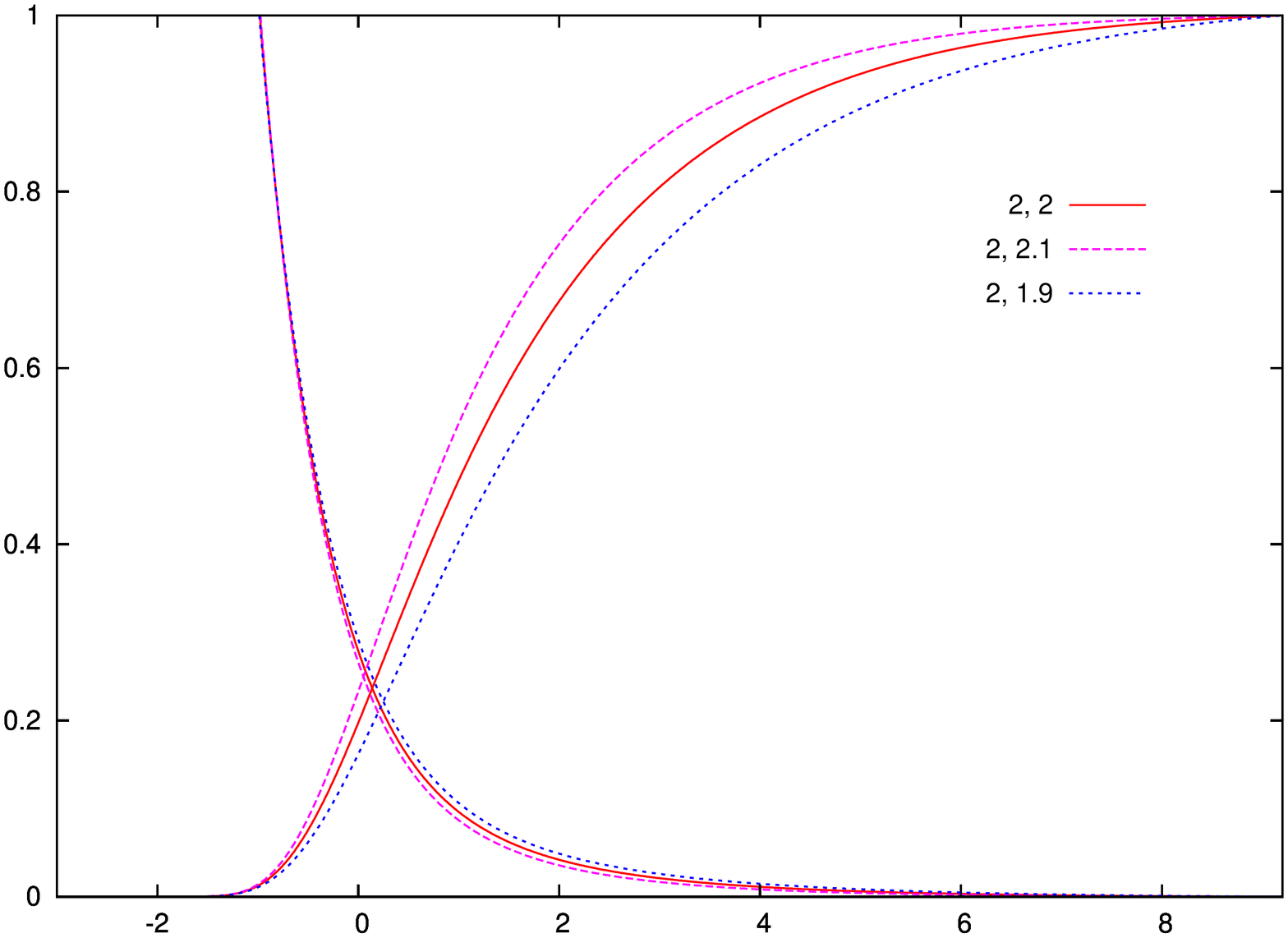}
\caption{Analytical solutions, $\frac{\theta}{\theta_{out}}$ against $x$
  (the ascending lines)
  and $\frac{\varphi}{2\pi}$ against $x$
  (the descending lines).
  The ($\beta_1$, $\beta_2$) values are given in the legend.
  $x$ is defined as $x=\ln(R/\Rw)$.
	}
\end{figure}

\section{Numerical solution}
We developed a finite differential code to solve the evolution of disc.
The state of the disc at each time point is represented
by the $\La$ value upon an uniform grid of $x$
(logarithemic grid of $R$).
The time differential of $\La$ are then evaluated,
and then the $\La$ value at next time point.
We used upstream differencing for the advective part in the equation.
The code is designed with flexibility to solve
various physical problems by adjusting the initial condition
and boundary condition.

The code can also be used in finding steady-state solution.
If the boundary condition is fixed to the desired setting,
and the evolution lasts long enough, in principle the disc
will always arrive at the wanted steady state solution.
However, the computational cost can be enormous,
due to the large time scale range involved in the system.
To ensure the solution reached the steady-state value,
the time $T$ of disc evolution much be at least several times larger than 
the viscosity time scale $T>R^2/\nu_1$.
On the other hand,
the maximum time step $\Delta t$
to keep the algorithm numerically stable
is determined by the time scale for \AM{} viscously diffuse over only one grid,
$2\Delta t< \frac{(R*\Delta x)^2}{\nu_1}$, where $\Delta x$ is the grid size.
These two conditions must hold
for the whole calculating region $R_{in}$ to $R_{out}$.
So the number of time steps needed are determined by
$\Bracket{\frac{R^2}{\nu_1}}_{max} /
  \Bracket{\frac{(R*\Delta x)^2}{\nu_1}}_{min} =
\Delta x^{-2} \Bracket{\frac{R_{out}}{R_{in}}}^{|2-\beta_1|}$.
As an example, supposing $\beta_1=0$, $R_{out}=10^4$, $R_{in}=10^{-4}$,
$\Delta_x=0.01$, we find $T>10^8\frac{R_0^2}{\nu_{10}}$,
$2\Delta t<10^{-12}\frac{R_0^2}{\nu_{10}}$,
so that the time steps needed are dozens of $10^{20}$, absolutely unaffordable.
Our way out of this difficulty is artificially
add a ``speeding up" factor $K(R)$
to the evolutionary equation eq.(\ref{Equation:AM}),
changing it to 
\begin{equation}
  \Ladot = K(R) {\cal F} (\La)
\end{equation}
where ${\cal F} (\La)$ is the time differential of $\La$
given in eq.(\ref{Equation:AM}).
This new equation leads to the right steady-state solution ${\cal F} (\La)=0$,
though its intermediate results
(the $\La$ values found before the disc get steady)
is physically meaningless.
We find $K(R)=R^{2-\beta_1}$ will make the equation converge stably and quickly.

In this work we set a uniform grid of $x$ from $x_{in}=-9.2$ to $x_{out}=9.2$
(corresponding to $R_{in}\approx10^{-4}$ and $R_{out}\approx10^4$,
the latter large enough to nearly infinity),
and the space resolution $\Delta x=0.01$.
We used a $\Bracket{\nu_1\La}'=\frac52\nu_1\La$
inner boundary condition,
by adding a ``ghost grid" at $x=x_{in}-\Delta x$,
and keep $\La(x)=e^{-(2.5-\beta_1)\Delta x} \La(x_{in})$,
in order to imitate a planary disc obeying $\nu_1\LaSca\propto R^{5/2}$
inside of the inner boundary.
At the outer boundary we set a fixed $\La(x_{out})$, with an inclination angle
to \BH{} spin axis (set as z axis).
The plane of z axis and $\La(x_{out})$ is set to be $xz$ plane.
So $\lVec(x_{out})=(\sin\theta_{out}, 0, \cos\theta_{out})$,
or $W(x_{out})=\sin\theta_{out}$.
We use the "solution B" (explained later)
as initial condition to save computational cost,
though the calculation can converge to steady state solution
from arbitrary initial condition.

In Fig.3, Fig.4 and Fig.5, the numerical solution
are shown and compared with solution A.
As an example, we show the results for $\beta_1=\beta_2=3/4$,
and the inclination angle at outer boundary to be
$\theta_{out}=\arcsin(0.01)$, $30\Degree$, $85\Degree$,
or equivalently, $W_\infty=0.01$, $0.5$, $0.9962$.
The numerical solution and analytical solution A coincides well
when the disc is only slightly misaligned $|W_\infty|\ll1$,
but when the inclination angle is large the two solution deviates strongly.
So we conclude that solution A is not appropriate for large
inclination angle.
In the plot of mass distribution,
we use $R^{\beta_1}\Sigma$ because analytical solutions predicts
$\Sigma\propto R^{-\beta_1}$ (similar as in planary disc).
The numerically calculated mass distribution differs from analytical solution
mainly in the vicinity of warping radius, showing a dip there.
This is natural because the warping there bring forth
additional \AM{} transfer,
so that the gas there falls faster than in the planary disc,
and thus cause a lower density there.

\begin{figure}
\center
\includegraphics[angle=-90,scale=0.30]{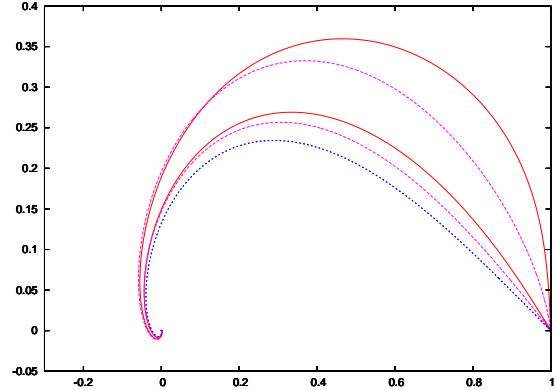}
\caption{
  Comparing the analytical and numerical solutions
  in the $l_y/W_\infty$ against $l_x/W_\infty$ plot.
  The $\beta$ values are $\beta_1=\beta_2=3/4$.
  From upside to downside, the lines are
  respectively:
  1. Numerical solution for $\theta_{out}=85\Degree$,
  2. Solution B for $\theta_{out}=85\Degree$,
  3. Numerical solution for $\theta_{out}=45\Degree$,
  4. Solution B for $\theta_{out}=45\Degree$,
  5. Solution A for all $\theta_{out}$ values,
    and also all the solutions for $\sin\theta_{out}=0.01$.
  All lines for solution A coincides,
  because solution A keeps $l_y/W_\infty$ and $l_x/W_\infty$
  constant for different $\theta_{out}$.
  All lines for $\sin\theta_{out}=0.01$ almost coincides,
    showing the error is negligible.
	In this and following two figures,
  we use solid lines for numerical solution,
	long dash lines for solution B,
	and short dash lines for solution A.
	}
\end{figure}

\begin{figure}
\center
\includegraphics[angle=-90,scale=0.30]{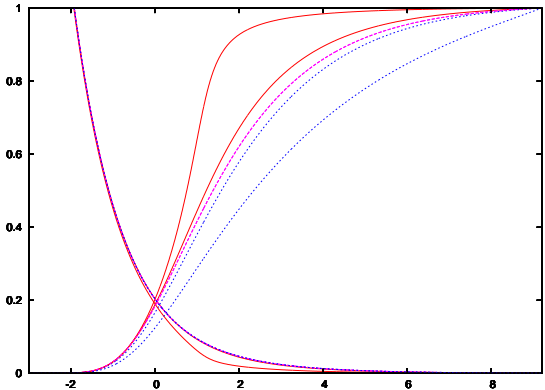}
\caption{
  Comparing the analytical and numerical solutions
  in the $\frac{\theta}{\theta_{out}}$ against $x$
  (the ascending lines)
  and $\frac{\varphi}{2\pi}$ against $x$
  (the descending lines) plots.
  The $\beta$ values are $\beta_1=\beta_2=3/4$.
  From upside to downside, the $\frac{\theta}{\theta_{out}}$ lines are
  respectively:
  1. Numerical solution for $\theta_{out}=85\Degree$,
  2. Numerical solution for $\theta_{out}=45\Degree$,
  3. Solution B for all $\theta_{out}$ values,
    and also all the solutions for $\sin\theta_{out}=0.01$.
  4. Solution A for $\theta_{out}=45\Degree$,
  5. Solution A for $\theta_{out}=85\Degree$,
  All lines for solution B coincides,
  because solution A keeps $l_y/W_\infty$ and $l_x/W_\infty$
  constant for different $\theta_{out}$.
  All lines for $\sin\theta_{out}=0.01$ almost coincides,
    showing the error is negligible.
  For $\frac{\varphi}{2\pi}$, the lower line is the numerical solution
    for $\theta_{out}=85\Degree$.
    The upper line is the analytical solutions
    (solution A and B give same $\varphi$, unvarying with $\theta_{out}$), 
    and the numerical solution for $\theta_{out}=45\Degree$
    and $\sin\theta_{out}=0.01$ also coincide with this.
	}
\end{figure}

\begin{figure}
\center
\includegraphics[angle=-90,scale=0.30]{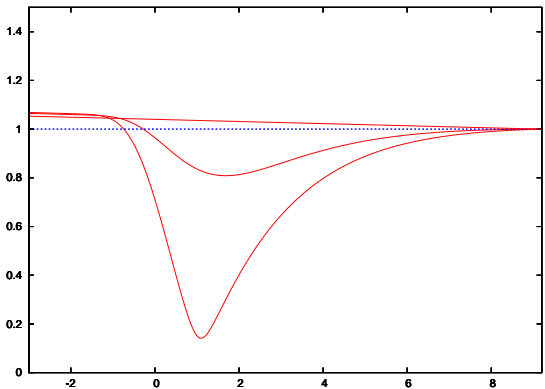}
\caption{
  Comparing the analytical and numerical solutions
  in the $R^{\beta_1}\Sigma$ against $x$ plot.
	The solid line: numerical solutions,
  for $\sin\theta_{out}=0.01$,
  $\theta_{out}=45\Degree$,
  and $\theta_{out}=85\Degree$,
  resepectively,
  from upside to downside.
	The short dash line: analytical solution.
	}
\end{figure}

\section{A new analytical solution for not so slightly misaligned disc}
To find a better analytical solution for more strongly misaligned disc,
we define another measure of misaligning $V=\theta(\cos\varphi+i\sin\varphi)$,
where $\theta$ and $\varphi$ are the inclination angle and azimuthal angle
of $\lVec$, respectively.
To the first order approximation of $\theta$, $W$ and $V$ equals,
$W=\sin\theta(\cos\varphi+i\sin\varphi)\approx V$.
So all the equations for $W$ in sec.3 also holds for $V$, hence we write
\begin{equation}\label{Equation:AnalyticalSolutionB}
V= f V_\infty = V_\infty \frac{2^{1-n}}{\Gamma(n)} s^n K_n(s)
\end{equation}
Hereafter we call this ``solution B",
and eq.(\ref{Equation:AnalyticalSolutionA}) ``solution A".
The two solutions is equivalent for slight misalignment,
but behave differently when extrapolated to
large inclination angle $\theta_{out}$.
When $\theta_{out}$ varies,
solution A keeps $\sin\theta/\sin\theta_{out}$ constant at each R,
while solution B keeps $\theta/\theta_{out}$ constant.
Thus solution A causes too quick a decreasing of $\theta$ at the outer disc,
while solution B gets rid of this backward.

Solution B is plotted in Fig.3 and Fig.4 to compare with solution A
and numerical solutions.
In the $W/W_\infty$ plot, solution A keeps unchanged
with different $\theta_{out}$,
while solution B predicts increasing $|W|/W_{\infty}$ with increasing $\theta_{out}$,
which is closer to the numerical results.
In the $\theta/\theta_{out}\sim x$ plot, solution B keeps unchanged,
while solution A predicts a decreasing $\theta/\theta_{out}$
with increasing $\theta_{out}$,
which contradicts the numerical results.
However, when $\theta_{out}$ are so large as $85\Degree$,
even solution B become very inaccurate.
On the other hand, for very small $\theta_{out}$,
the two analytical solutions are equivalent and both very accurate.

\section{Conclusions}
We generalized MPT07's analytical solution of warped accretion discs
to the situation
that the power law index of the two \VC{s} is not necessarily equal
(solution A).
We then proposed a new analytical solution (solution B),
which is supposed to be more accurate then solution A.
We also presented the numerical solutions
of the dynamical equations for warped disc.
Our comparison between the two analytical solutions and the numerical results
show that solution B is indeed better and is recommendable for 
moderately or slightly misaligned disc.
For extremely misaligned disc, only numerical solution is appropriate.
As for the situation in NGC4258, M08's fitting suggested a large
inclination angle, so that numerical solution is needed
for more accurate fitting.

\section*{Acknowledgments}
This work was supported in part by the Natural Science Foundation of China
(grants 10773024, 10833002, 10821302, and 10825314),
Bairen Program of Chinese Academy of Sciences,
and the National Basic Research Program of China (973 Program 2009CB824800).

\bibliography{head,WarpedDisk}
\end{document}